Challenges and opportunities in visual interpretation of Big Data

Gourab Mitra

University at Buffalo

Author Note

Gourab Mitra, Department of Computer Science and Engineering, University at Buffalo

Correspondence regarding this article should be addressed to Gourab Mitra, Online Data Interaction Lab, Department of Computer Science, University at Buffalo, Amherst NY 14260.

Contact: gourabmi@buffalo.edu

**Abstract**

We live in a world where data generation is omnipresent. Innovations in computer hardware in the last few decades coupled with increasingly reliable connectivity among them have fueled this phenomenon. We are constantly creating and consuming data across digital devices of varying form factors. Leveraging huge quantities of data involves making interpretations from it. However, interpreting data is still a difficult task. We need data analysts to help make decisions. These experts apply their domain knowledge, understanding of the problem space and numerical analysis to draw inferences from the data in order to support decision making. Existing tools and techniques for interference serve users making decisions with hard constraints. Consumer systems are often built to support exploratory data analysis in mind rather than sense making.

*Keywords*: Data Visualization, Big Data

Challenges and opportunities in visual interpretation of Big Data

We are in the middle of a big data revolution. We are generating more data than ever ; both in terms of quantity and type. But it is important to distinguish between the marketing buzz around big data in order to understand what it really means. Let us look at big data from the lens of the attributes attached to the classical definition of data. [2]

1. Volume – The most intuitive characteristic about data is the amount of data itself. In fact, it is the sheer amount of data that we generate and process these days that poses challenges to interpreting it.

2. Velocity - Velocity refers to the idea of the amount of data flowing through an interface in unit time.

3. Variety - Traditional data formats were relatively well defined by a data schema and changed slowly. In contrast, modern data formats change at a dizzying rate. This is referred as variety in data.

4. Value - The value of data varies significantly. The challenge in drawing insights from data is identifying what is valuable and then transforming and extracting the data for analysis.

Our working definition of big data in this paper would be [3] a *technology to process high-volume, high-velocity, high-variety data or data-sets to extract intended data value and ensure high veracity of original data and obtained information that demand cost-effective, innovative forms of data and information processing (analytics) for enhanced insight, decision making, and processes control.*

There exists many different tools and techniques for data analysts; mainly drawn from various disciplines like statistics and computer science. Visualization methods concern the design of graphical representation to represent the innumerate amount of the analytical results as diagrams, tables and images. Data visualizations inherently end up summarizing the data that they represent. Research shows that 98 % of the most effective companies working with Big Data are presenting results of the analysis via visualization. [4] Visualizations enable rapid exploration of data and identify relationships in it.

Information Visualization [5] emerged as a branch of the Human-Computer Interaction field in the end of 1980s. It utilizes graphics to assist people in comprehending and interpreting data. As it helps to form mental models of the data, for humans it is easier to reveal specific features and patterns of the obtained information.

Visual Analytics [6] combines visualization and data analysis. It has absorbed features of Information Visualization as well as Science Visualization. The main difference from other fields is the development and provision of visualization technologies and tools.

Efficient visualization tools should consider cognitive and perceptual properties of the human brain. Visualization aims to improve the clarity and aesthetic appeal of the displayed information. It allows a person to understand large amount of data and interact with it. Significant purposes of Big Data visual representation are: to identify hidden patterns or anomalies in data; to increase flexibility while searching of certain values; to compare various units in order to obtain relative difference in quantities; to enable real-time human interaction.

Whether a new product review on Amazon, a social network posting, minute-by-minute investment data, or a shared document in the cloud; users create, consume, and disseminate an enormous amount of valuable information. Interpreting such rich sources of data can be a daunting task. While sources of 'big data' have evolved and expanded over time, our methods of inference have not kept pace. The process of extracting meaning from these rich datasets is dependent upon identifying 'useful' examples effectively. Human perceptual capabilities are not sufficient to embrace large amounts of data. *Olshanniova et. al.* describe this issue using the following venn-diagram analogy [8].

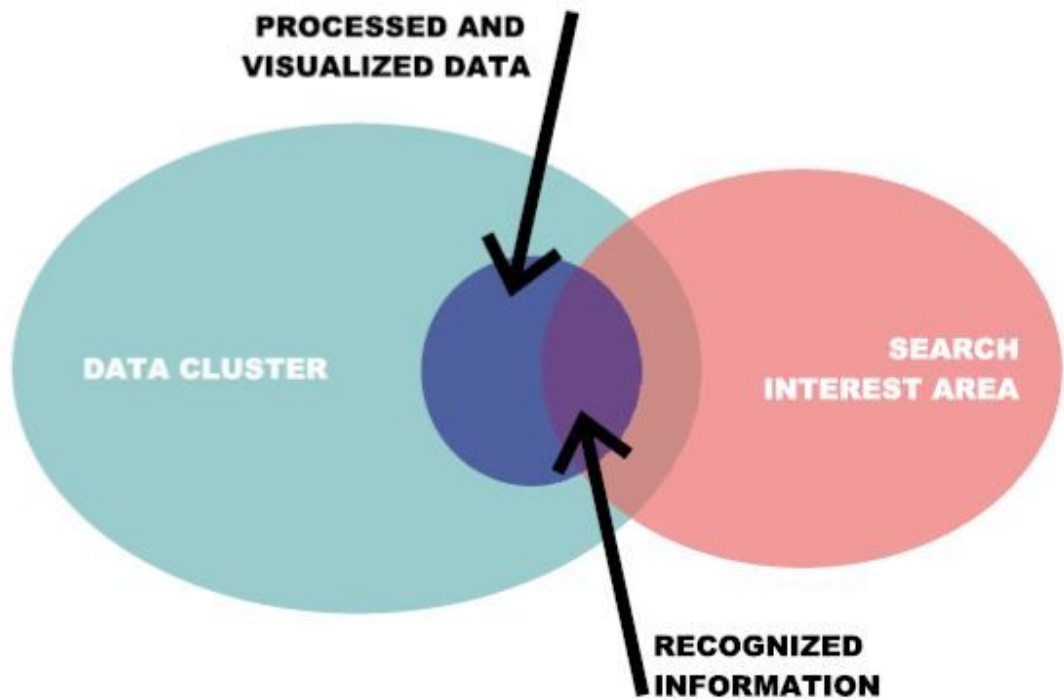

Figure 1. Human Perception Capability Issue [8]

Exploratory data analysis enables us to crunch through large volumes of data to develop a frame that describes it. Examining many points can be wasteful because the analyst needs to shift attention repeatedly. Digital visualizations are often designed to either support exploration or enrichment. Let's consider an example.

Search interface in Google Scholar is effective in returning useful queries when during exploration of published research literature. Users have a much harder time organizing and managing the results, necessitating efforts that help users curate and rediscover knowledge they develop during course of searching. [7] As the dimensions of data keep on increasing, filters with discrete values or range and similar hard constraints become insufficient to model the information need.

## Towards effective visualizations

*Rzeszotarski* describes an approach for generating data visualizations to address these issues[1]. Users constantly interpret data by taking small steps that generate insight while exploring the data.These steps include developing hypotheses, and testing them through experimentation and observation before finally returning to exploration. A key challenge is to help users develop a consistent model when they explore data. Actions that cause changes in the visualization of data (for example  applying filter to remove points and instantly relocate others as axis scales adjust) tend to disrupt user's perception of the scene and also affect recall and require more attention [10]. Moreover, users make incremental progress while analysing. More often than not, most data exploration systems do not reflect a trace of their past action which have lead them to the current state. Users may not remember the particular steps that lead to their current position on data.  Another challenge while data scales up is to help users develop models for information that may surpass their working memory limits.As data scales up, another challenge is to help users model information that may exceed their working memory limits. For instance, a user attempting to track a relationship that spans three or four dimensions may have to compare between three or four different charts to find it, due to interface restrictions .This could get taxing and difficult to manage , even for an expert.

Going by the fact that sensemaking performance is influenced by motivation and efficacy [11], interfaces should encourage exploration and extensive data digging. Complex interfaces may require long periods of training and might be overwhelming and intimidating for new comers [12]. Current research provides an insight into  not only delivering easy and motivating interfaces, but also solving the problems of enhancing recall and modeling mentioned in the previous paragraph. Users are already well trained for interaction with objects in the real world. By designing exploratory visualization tools that use physics and naturalism to match users' real life experiences, fluidly move between analysis steps (as real objects do), and utilize touch modalities that bind users and their interface actions [13], exploratory performance can be improved.

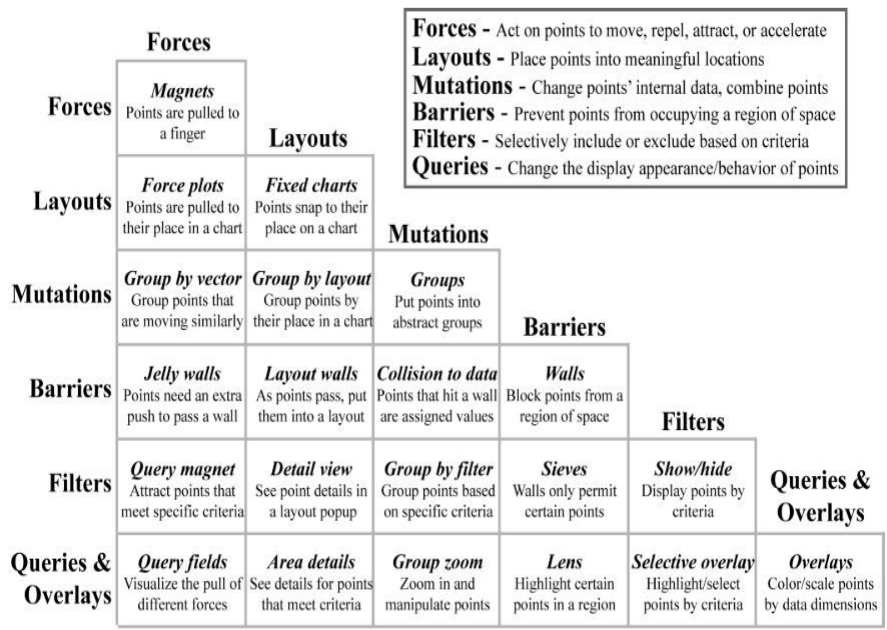

Figure 2. Visualization generation framework [1]

## Generative framework for big data visualizations

*Rzeszotarski* describes a naturalistic or physics-based visualization affordance as an interaction or visual affordance that resembles a real world physical phenomenon. This is described in Figure 2. For example, Dust & Magnet [9] employs a naturalistic magnet affordance for setting up data in its data analysis process. Just as iron filings get attracted to magnet, data are attracted with varying force to Dust & Magnet's canonical representations. The distinction between naturalism and realism is put to use here. The Dust & Magnet magnets do not resemble magnets. and their magnetic effect does not decay with a lifelike inverse square. The visualization affordance resembles the real world. It is effective metaphor, but does not necessarily mimic the real world. While this is largely up to the interaction/visualization designer, the general design philosophy towards naturalistic visualizations has been to use pure mimicry strategically and sparingly.

Physics-based visualization affordances make use of the inherent expertise users have based on their experiences in the everyday world in order to help them develop an understanding of data. These techniques are different from traditional visualization approaches, and, in light of users desire at times for more familiar controls, may work well in concert. However, because they are different from traditional approaches, it is not always easy or intuitive to create new interactions. In considering and developing naturalistic affordances, Rzeszotarski's framework outlines several different modes of interaction with data entities.

It is composed of the following ideas:

- Physical points represent data. They  have associated physical properties that correspond to their values in different dimensions

- Data occupy a logical sandbox that contains them. It allows for interaction. Interactions with the sandbox change the physical arrangement of the data in the space, and leave traces of their activities.

- The user can employ forces to act on physical points either independently or as a result of their unique data. For instance, a magnet may repulse points with low values in a particular dimension.

- The user may use layout tools to force points into strict, meaningful locations, breaking with the physics metaphor when necessary (such as when allowing points to pass over or under others to avoid being trapped).

- The user can mutate points, for example combining multiple points into one group so as to observe more points at once. It can be used infer larger trends.

- The user can impose conditional barriers that block or selectively block points based on criteria.

- The user can filter out points to help avoid overload or choose only a small subset of interest.

- The user may use queries and overlays to change the appearance or behavior of points on the screen.

**Conclusion**

This framework could be used in many different ways. This could be used to generated interactive visualizations which can function using a keyboard, mouse or multi-touch, though multi-touch offers the greatest opportunity for naturalism and closeness to the user [13] Furthermore, these concepts can be combined to generate a much richer set of potential physics based affordances. Mixing different primitives together provokes new ways to augment data visualizations with physics.

<div align="center">**References**</div>